\documentstyle[12pt,twoside,fleqn,espcrc1]{article}

\newcommand{\AmS}{{\protect\the\textfont2
  A\kern-.1667em\lower.5ex\hbox{M}\kern-.125emS}}

\title{Can a flavour-conserving treatment improve things ?}

\author{E. Mendel\address{ FB Physik,
 Carl von Ossietzky Universit\"at Oldenburg,\\
         26111 Oldenburg, Germany}\thanks{Talk presented 
at the Intl.
Workshop on QCD at Finite Baryon Density in Bielefeld, April 98.}}

\begin{document}
% typeset front matter
\maketitle

\begin{abstract}
  In this work I would like to present some ideas on how to improve
on the gauge sector in our lattice simulations at finite baryon density.
The long standing problem, that we obtain an onset in thermodynamic
quantities at a much smaller chemical potential than expected, could
be related to an unphysical proliferation of flavours due to hard
gluons close to the Brillouin edges. These hard gluons produce 
flavour non-conserving vertices to the fermion sector. They also
produce excessive number of small instantons due to lattice
dislocations. Both unphysical effects could increase the propagation
in (di)-quarks to give the early onset in $\mu$. Thus we will present
here a modified action that avoids large  fields close to the
lattice cutoff. Some of these ideas have been tested for $SU(2)$
and are being implemented for $SU(3)$.    
\end{abstract}

\section{Introduction}

   There has been the long standing problem, since the introduction
of a chemical potential $\mu$ coupled to the baryon density 
\cite{Kog,Kar}, that the onset of thermodynamic quantities happens
in our lattice simulations \cite{Bar,Men3,Kle} at a much smaller $\mu$  
than at the expected $\mu = m_N /3$. The baryon density and the
chiral condensate start changing from their vacuum values at $\mu$
of the order of  or earlier than  $m_{\pi} /2$, as if controlled by a
Goldstone mode with net Baryon number, which is clearly unphysical
in the low Temperature normal phase. 

  The community had the hope that including properly the fermion
determinant, which for non-zero $\mu$ is complex and therefore
very difficult to implement with lattice methods, will by itself
solve this problem. The careful work by the Barbour group  in ref.\cite{Bar2}
and the follow up presented by I. Barbour and S. Morrison at this
Workshop, seem to show a pinching at a higher $\mu$ but the onset still 
seems to be around $m_{\pi} /2$.

  Several years ago I described another possible problem, 
on top of the fermion determinant one, that could be producing the 
early onset in our lattice simulations. 
 As is well known, for Kogut Susskind fermions the flavour
number is not conserved along their propagation due to hard gluons that
can make jump the fermion propagator pole to other corners of
the Brillouin zone, thus changing its flavour. This flavour changing
would produce a proliferation of flavours even for the valence quarks,
forming a much tighter bound nuclear matter than in nature\cite{Men3,Men4}.
 As we heard on various talks by F. Wilczeck, E. Shuryak and others at 
this workshop, diquark condensates
could be produced easier the  more flavour species, which could drive
the observed early onset due to the flavour proliferation. Furthermore,
these hard gluons, with momenta $p \approx \pi /2a$ or higher, produce
easily lattice dislocations that give many unphysical
 ``small instantons''\cite{Men5}.
 Again E. Shuryak in his model of quark propagating among instantons, gets
increased propagation with a higher instanton density.

  We see then, that from several points of view it seems important
to find a way to suppress the gluons with momenta of the order of
the lattice cutoff.  We could regain flavour conservation and
suppress unphysical mini instantons. Both unwanted effects tend to lower
the $\mu$ onset value from the physical one.

  In order to achieve such a hard gluon suppression I have been experimenting,
first for $SU(2)$, with an ``improved'' gauge action which has the same naive
continuum limit as the usual one in the field strengths, but diverges for
plaquette phases of the order of the lattice cutoff. Even though this suppression
method is gauge invariant, it is important to use a Metropolis updating
scheme with small changes in the link phases, in order to guarantee that we
stay locally without spurious dislocations. In this way we can hope to
suppress small unphysical instantons.  The physical gluon momentum
transmitted to the fermions, given in the continuum by the Poynting vector,
being essentially in the nonabelian case shaped as a ``chair'' formed with two 
contiguous plaquettes, could then also be kept low so as to conserve flavour.

  With L. Polley we had also tried out a new kind of lattice 
fermions \cite{Pol,Men6}, 
the \mbox{t-asymmetric} ones, where one can reduce to 2 flavours from the start
(in contrast to the 4 staggered ones), but have several technical problems
to surmount, related with regaining the $O(4)$ symmetry, in order to calculate
the hadron masses and compare them with the onset. This will be treated in
the next section.

   In sec. 3 we will describe the new gauge action, mainly for $SU(2)$, 
and discuss its implementation for $SU(3)$. The ultimative test will be
to generate configurations generated with such an improved gauge action, to
then compute the baryon density or chiral condensate for the full theory.

\section{t-asymmetric fermions}
  
  An earlier attempt in order to reduce the number of flavours, was 
based on a new kind of fermions, also called t-asymmetric fermions,
that don't allow more than 2 flavours in contrast to the 4 flavours
for staggered ones \cite{Pol,Men6}. This fermions are obtained by just
taking a one-sided time derivative, which still produces an hermitian
hamiltonian and eventually a positive transfer matrix 
(for $a_\tau < a_s / \sqrt{3+m_q^2}$). Consequently, the number of
flavours gets reduced to just 2. These fermions are related to
Susskind's hamiltonian formalism ones, via a unitary transformation
\cite{Suss,Banks} and a related form is also being investigated by
the Heidelberg group \cite{Sta}.

  The details of this formulation can be found in \cite{Pol,Men6}, with
the result that after a diagonalization in spinor space like the
usual Kawamoto-Smit one,
\begin{equation} \label{2}
\psi(x) = \alpha_1^{x_1} \alpha_2^{x_2} \alpha_3^{x_3}
   \chi(x) \quad\mbox{with}\quad
    \alpha_k = i \gamma_4 \gamma_k
\end{equation}
we get,
\begin{equation} \label{3}
K_{xy} = m \Gamma_4(x) \delta_{xy}
 + {\textstyle\frac{1}{a_{\tau}}}
  \left( e^{\mu} U_{x,4}\, \delta_{x,y-\hat{4}}
 - \delta_{x,y}\right)
 -  {\textstyle\frac{i}{2 a_{\sigma}}} ~ \sum_k
\Gamma_k(x) \left( U_{x,k} \,\delta_{x,y-\hat{k}}
  -  U^{\dag}_{x-\hat{k},k}\, \delta_{x,y+\hat{k}}\right)
\end{equation} 
with the usual:
$\Gamma_{\nu}(x) = (-)^{x_1+\cdots + x_{\nu-1}} $, after
having thinned out to the first diagonal component. This
fermionic action $ \bar{\chi} K  \chi $ has then just 2
flavours. Susskind showed that there is a {\em discrete} version
of chiral invariance left over, preventing mass counterterms in
the interacting case, but not guaranteeing ``Goldstone behaviour'' for
the pions. One could still get light pions, like in Aoki's proposal
for Wilson fermions \cite{Aoki}.

  We had already done first simulations with the baryon density
for these fermions,
$ \langle J_4 \rangle = \langle e^{\mu} \,\bar{\chi}_x
  \, U_{x,4} \, \chi_{x+\hat{4}} \rangle $
and the chiral condensate, $\langle \bar{\chi} \chi \rangle$,
as function of
$\mu$. The operators have been calculated with the solved
pseudofermion method \cite{Men3}.
The curves for $\langle J \rangle$ are shown in ref. \cite{Men6} (Fig. 1)
and show a clear onset for two  $m_q=.01,.04$ at the same $\mu \approx 0.2$. 
This result could be deceiving though as it is not clear
how  $m_\pi$ scales with $m_q$ for these fermions, as already mentioned 
above. In order to really
check if we are getting better results, we have to compute for these
fermions the $\pi$ and $N$ masses. For this, considering that we need
a finer lattice in the $\tau$ direction for positivity and a fine
$a_\tau$ for charge conjugation symmetry in the propagators, we
have to scale the couplings to regain the $O(4)$ symmetry. Once this is
done, one can attempt to calculate the masses and the baryon density
for this $O(4)$  symmetric action. This has proven to be a formidable
task and presently I consider more promissory to improve on the
gauge sector.

\section{Improvements on the gauge action}

  The other way out, in order to prevent flavour changing is to suppress the
appearance of hard gluons with momenta $p > \pi /2a$. Without these hard 
gluons the quarks cannot change their flavour by jumping to other corners
of the Brillouin zone. As mentioned, these hard gluons can also cause
unphysical mini-instantons that could enhance the quark propagation
in Shuryak's scheme, also affecting the $\mu$ onset.

  From my investigations in $SU(2)$ with instantons, in order to try to
suppress mini-instantons produced by lattice dislocations, one can do so
by modifying the action for plaquettes with a large phase. Instead of
the usual gauge action:   
\begin{equation} \label{4}
S = \beta \  \sum_{\Box}  \  (1- 1/N \  {\rm Re \ tr} ( U_{\Box})),
\end{equation} 
which has a maximum plaquette action of $2 \beta $  for a phase $\theta_{\Box}=\pi$, 
one can take an  improved action of the form:              
\begin{equation} \label{5}
S = \beta \   \sum_{\Box}  \ 2/\pi \ \tan(\  \pi/2 \ (1- 1/N \ {\rm Re \ tr} ( U_{\Box}))),
\end{equation} 
which has the same naive continuum limit for small $a$, but diverges for plaquettes with
phases close to the cuttof. One can easily device functions which start the same and
grow even faster for larger plaquette phases. The point being that at relatively moderate
$\beta$'s one can suppress plaquette phases larger than $\theta_{\Box} \approx \pi/2$ and
therefore make sure that the effective ${\rm  tr}( F_{\mu \nu}^2)$ stays away from the
cutoff. ( For the usual gauge field action in $SU(3)$, for example at $\beta=6.0$,
around 1/3 of the plaquettes still have a phase $\theta_{\Box}> \pi/2$ !).   
 The actual gauge invariant definition for the gluon momentum, resembling the
Poynting vector (${\rm tr}{\bf E \wedge B}$) in the continuum, involves  pairs of 
orthogonal plaquettes like a ``chair''. If each of the plaquette phases is kept small
also the combined one should be constrainable to get $p < \pi /2a$.

  In order to avoid large phases in the gauge links, which produce dislocations
carring a topological charge \cite{Men3}, it is convenient to use the Metropolis
algorithm for the link update, with small changes in the $SU(N)$ matrix on each iteration.
In combination with an improved action as introduced above, one can obtain then fairly
``smooth'' gauge configurations, starting from the identity one. There is a problem
though, related with a random walk like drifting of the links at a vertex 
corresponding to local gauge transformations, until we get
large link phases close to $\pi$ that can produce dislocations. This can be reduced by   
gauge transformations that try to keep all the link phases low. The trick I used in
$SU(2)$ is to check the phase $\phi$ of the new link 
$U = \exp (i \ \phi \  \hat{{\boldmath \phi}} \cdot  {\boldmath \sigma} )$ and if it is 
larger than some phase,
lets say $\pi/6$, make a local gauge transformation that shrinks that link's phase 
to $\pi/8$ keeping its $SU(2)$ direction, while applying the inverse element to the 
other links at that vertex.
  These local gauge transformations succeed in keeping almost all $\phi$ smaller than
 $\pi/6$, thus suppressing almost all dislocations. 

  In this way I have been able to constraint the total topological charge to values
\mbox{$Q_V < |\pm 1|$} in periodic lattices (the field-theoretic definition does not give
exactly integers), for fairly large $SU(2)$ lattices ($12^3 \times 24$) 
with $\beta=2.5$. For these lattices also the charge  $Q_\upsilon$ with $\upsilon=V/2$
half the volume is also smaller than $ |\pm 1|$ showing no $I - \bar{I}$ pairs.        

  One could also use more conventional action improvements in order to avoid hard
gluons, by including terms in the action proportional to the Poynting operator
(more generally the energy-momentum tensor $\Theta_{\mu \nu}$) that suppress
large fields. In other words, include ``chair'' diagram terms in the action.
   
  Another interesting alternative for an improved action in order to
reduce the flavour breaking, was presented by Laga\"{e} and Sinclair \cite{Lag}.
 They smear the links by covariant displacements in all directions, in order
to {\em project} the momenta to the central pole region in the Brillouin zone.
 In momentum space  in the continuum they project the fields 
$ A_{\mu}(k) \rightarrow 1/16 \ \prod_{i=1..4}(1+\cos k_i ) \ A_{\mu}(k)$, which
in coordinate space corresponds to the smearing  
$A_{\mu}(x) \rightarrow 1/256 \ \prod_{i=1..4}(2+ D_i + D_{-i} ) \ A_{\mu}(x)$.
They report a large reduction in the flavour symmetry violations for the 
(non)-Goldstone $\pi$ splitting of masses. Up to now the only introduce this
projected gauge fields in the minimal coupling terms in their action and it
would be interesting to see what happens with the topology if they consider
such improvement in the gauge action. 

\newpage
\section{Conclusions}

  The inclusion of the fermion determinant could  be not the only ingredient 
needed in order to shift the onset chemical potential to physical values close 
to the Nucleon mass.
Our lattice simulations at finite $\mu$ have been done  with an unimproved gauge
sector. This produces several unwanted effects, like the proliferation of
flavours even for valence quarks due to the hard gluons that can change the
flavour along the quark propagation. Also the topological sector is distorted
due to unphysical mini-instantons appearing due to non smooth gauge fields.
  
  Both unwanted effects should be suppressed with an improvement in the gauge
action. I introduced a simple to implement improvement that should help avoid
hard gluons and dislocations. These ideas have been tested for $SU(2)$ with
encouraging results and could also be implemented for $SU(3)$. In the Metropolis
algorithm we cannot use anymore the trick to have the sum of staples precalculated
as the action is not linear anymore, but the rest is very simple to implement.

  Once we have the gauge sector under control, with no hard gluons close to the
cutoff and no unphysical mini-instantons, we could again calculate the baryon
density operator or the chiral condensate to find its $\mu $ dependence. This
gauge updating could also be integrated in the unquenched code to redo Barbour's
group method to include the determinant.

  I would like to thank L. Polley for several interesting conversations on
this gauge field improvement.

\end{document}